# Stable room-temperature multiferroic skyrmions in lithium niobate with enhanced Pockels effect

Yalong Yu[1], Bo Xiong[1], Siqi Wu[2], Yekai Ren[1], Nuo Chen[1], Qingjiao Mi[1], Kangping Lou[1], Rui Wang[1], and Tao Chu[1†]

[1]College of Information Science and Electronic Engineering, Zhejiang University, Hangzhou, 310058, P. R. China

[2]School of Physics, Zhejiang University, Hangzhou 310058, P. R. China

[†]E-mail: chutao@zju.edu.cn;

## Abstract

**Lithium Niobate (LN) is a ferroelectric material with exceptional electrical characteristics, including high piezoelectricity, high Pockels effect, etc[1]. These properties make it a promising platform for numerous fields such as high-speed communication[2], optical computation[3], and quantum information processing[4]. Besides these, the introduction of magnetic structures to LN holds significant potential to achieve magnetoelectric coupling, which can be applied in magnetic memory and data-processing devices with high efficiency[5-7]. Here, for the first time, we observe a special topological magnetic structure called magnetic skyrmion[8] in LN (SK-LN) by the combination of magnetic field annealing and rapid annealing processes. Compared to the magnetic skyrmions reported in magnetic systems, SK-LN exhibit exceptionally high stability. It's stable at room temperature in the absence of external fields, displaying a high critical temperature over 673 K and a critical field over 0.7 T. Additionally, the center of the magnetic vortex exhibits spontaneous ferroelectric polarization, indicating its multiferroic characteristic. With the**

**excitation of these multiferroic skyrmions, the modulation efficiency of the electro-optical (EO) modulator fabricated on thin film lithium niobate on insulator (LNOI) wafer was found to be enhanced from 1.98 V·cm to 0.63 V·cm, which is the highest value as far as we know. It is considered that the multiferroic skyrmions significantly enhance the Pockels coefficient of LN to 101 pm/V, nearly three times the result (32pm/V) reported previously. These results reveal the first observation of room-temperature magnetic skyrmion in ferroelectric materials. Given its exceptional stability, Pockel effect enhancement ability, and potential for magnetoelectric coupling capability, our findings will have a significant impact in the fields of condensed physics and optoelectronics.**

## Introduction

As an artificial ferroelectric material, LN has been extensively researched in past decades and is considered to be one of the ideal platforms for high-speed communication, quantum computing, and photonic accelerators[2,4,9]. Up to now, LN has been reported to exhibit various electrical properties, including piezoelectricity, thermoelectricity, the Pockels effect, the Kerr effect, and more[1]. Piezoelectricity and thermoelectricity enable the conversion of phonon energy into electric energy or vice versa. The Pockels effect and Kerr effect enable the modulation of light signals in LN via an applied electric field. Therefore, LN can be utilized in various devices such as acousto-optic modulators, EO modulators, radio-frequency filters and piezoelectric sensors[10-13]. In recent years, photonic integrated circuits (PICs) based on thin film lithium niobate (TFLN) also have gained significant attention for the potential to achieve high-speed communication over THz in theory due to the high intrinsic Pockels coefficient of LN (32 pm/V)[11,14-17]. It is expected to replace the PICs based on silicon where the conversion of electric signals is achieved through the plasma dispersion effect, facing significant challenges above 100 GHz for its physical limit.

Besides ferroelectricity, magnetism, as one of the long-range ferroic orders, has also found extensive applications in data storage, energy conversion, and signal processing. The investigation of the interaction between diverse electrical properties and magnetism in LN may facilitate the exploration of magnetoelectric coupling in condensed matter physics and help advance the realization of next-generation magnetic storage with electric field-driven motion[5-7]. Additionally, due to the potential magnetoelectric coupling, it is possible to regulate the initial electrical properties of LN through the magnetic structure within it, which may help improve its initial electrical properties. The magnetic properties of original LN do not offer significant advantages as most LN crystals exhibit paramagnetism, with only a minimal manifestation of weak ferromagnetism attributed to lattice defects[18]. Currently, there are mainly two methods for introducing magnetism into LN: increasing lattice defects or doping it with magnetic elements[19,20], both of which can induce significant ferromagnetism in LN. However, there is currently no evidence demonstrating the direct coupling between the newly induced ferromagnetic properties and the inherent ferroelectric properties of LN by these means, indicating limited applications of these special LN systems. On the other hand, introduced defects or magnetic dopings will compromise the original ferroelectricity of LN[20]. Therefore, generating magnetic structures with magnetoelectric coupling potential in LN remains a significant research challenge worth exploring.

In this work, we report the first observation of room-temperature magnetic skyrmions (a nanoscale magnetic whirl-like spin texture with topological protection) excited in LN by means of magnetic field annealing (see Supplement.Fig 3 for detailed procedures), a process that has been adapted widely in the modification of soft magnetic ferrite materials[21]. These skyrmions exhibit stability in the absence of external fields, displaying a compact size of approximately 30 nm in diameter and a high critical temperature of over 673 K. Furthermore, through a rapid annealing process, these magnetic skyrmions undergo a transformation into multiferroic skyrmions accompanied by the emergence of spontaneous ferroelectric polarization. With the excitation of the dense multiferroic skyrmions in LN, we observed an enhancement in the modulation efficiency of the EO modulator fabricated on the TFLN platform, resulting in a reduction of the half-wave voltage-length product from 1.98 V·cm to 0.63 V·cm. The Pockels coefficient of LN with multiferroic skyrmions can be evaluated as 101 pm/V, three times of the results of 32 pm/V in LN as reported before.

## Identification of skyrmions

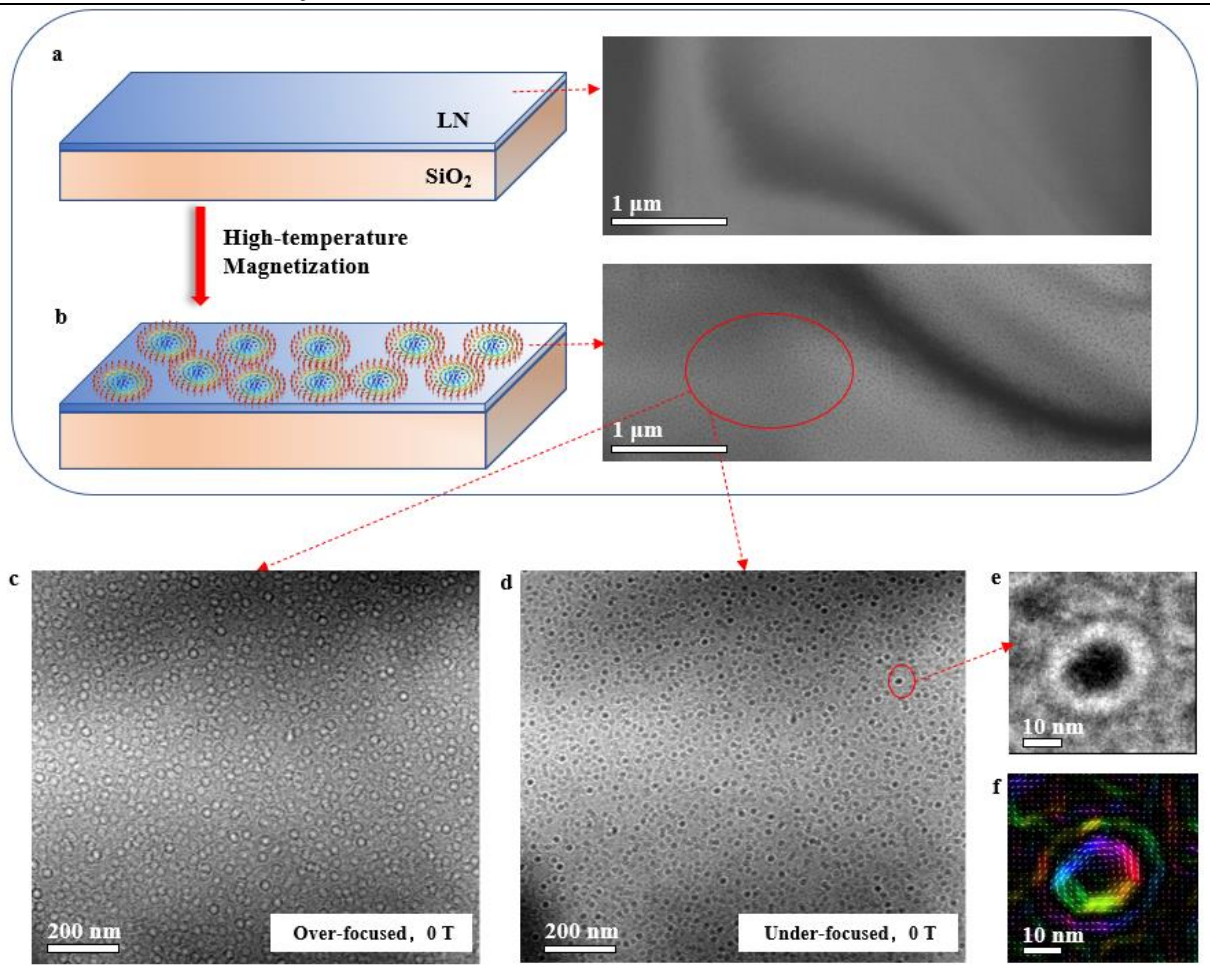

**Fig. 1** | **Room-temperature skymions in LN**. **a,b**, The under-focused LTEM images of LN before and after magnetic field annealing respectively. **c,d**, The over-focused and under-focused LTEM images of (**b**) in a narrow field of view. **e**, A single skyrmion shown in (**d**). **f**, The magnetic induction map of (**e**), in which the color and white arrows indicate the magnetization direction at each point.

Typically, magnetic skyrmion is a stable whirl-like spin texture, originally predicted by Tony Skyrme in the 1960s and first observed experimentally in real space in 2009[8,22]. This unique topological magnetic chiral structure has garnered significant attention from researchers for its highly promising applications in the field of spintronic devices such as racetrack memory, Hall sensors, logic gates, nano oscillators, and neural network caculation [23,24]. Over the past decade, this magnetic texture has been extensively investigated and observed in various ferromagnetic families, including B20 compounds, 2D systems, heavy-metal/ferromagnet multilayers[25-28], etc. In this study, we find that magnetic skyrmions can be excited in LN, a novel ferroelectric environment different from these aforementioned systems. Here, these magnetic textures were observed in real space using the Lorentz transmission electron microscopy (LTEM) technique, enabling the identification of their species and determination of their sizes[22,29,30].

Fig 1 presents the LTEM images of LN film without an external magnetic field at a temperature of 300 K. Fig 1 (a) depict the under-focused LTEM images of initial LN film substrated with $SiO_2$, from which no magnetic structure can be observed. The result of LN film after magnetic field annealing is shown in Fig 1 (b), revealing the emergence of a distinct substructure composed of vortex patterns.To clarify the type of these patterns, the LTEM images of LN films were further examined under both on-focused and under-focused conditions with the magnitude of deviation in focal length as 90 μm, using a narrower field of view, shown in Fig 1 (c,d). The vortex patterns exhibited opposite contrasts in the cases of under-focused and over-focused, providing evidence for their magnetic skyrmion nature[22,29,30,31]. The LTEM image of a single skyrmion at an under-focused state and its magnetic moment morphology processed by the Transport of Intensity Equation (TIE)[32] algorithm is shown in Fig 1 (e, f). It can be confirmed that the skyrmion in LN has a typical size below 30 nm, and its skyrmion number Q is equal to 1.

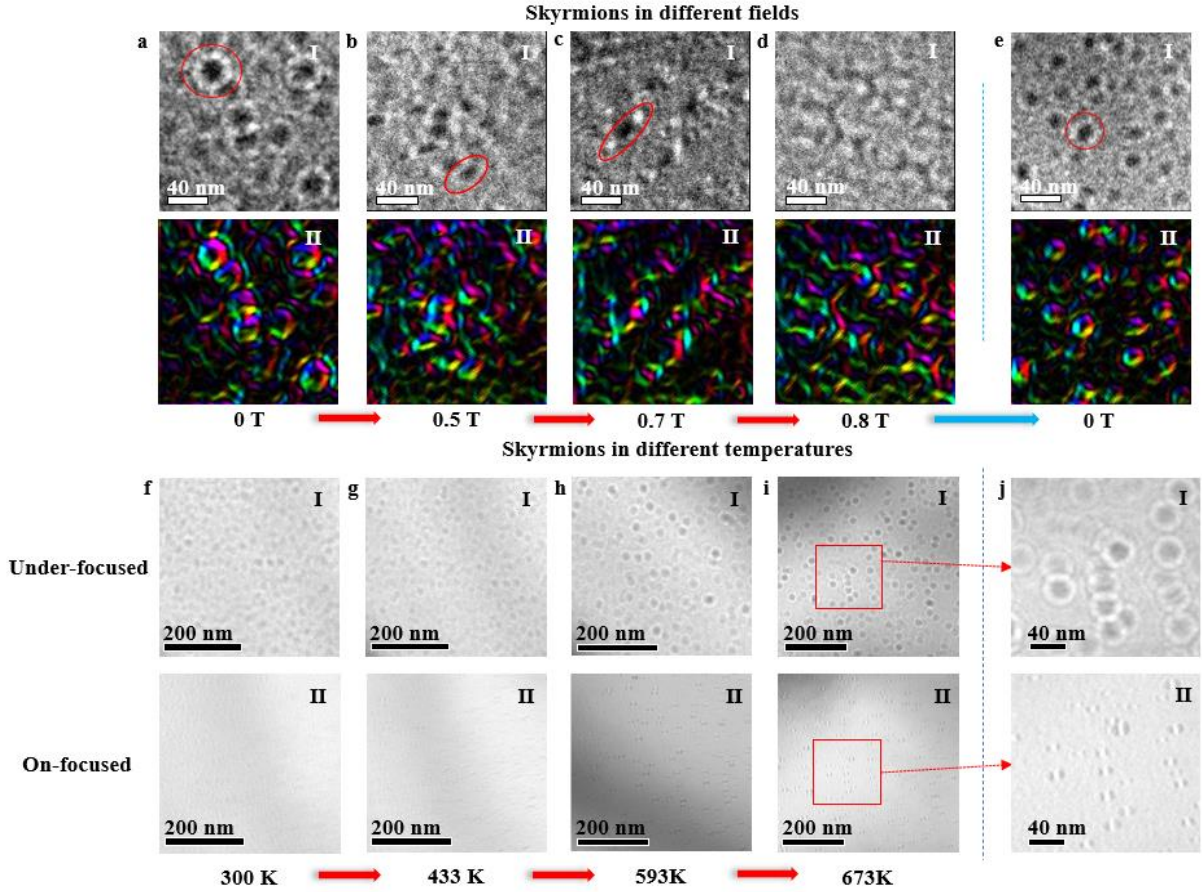

**Fig. 2 | Skyrmions under different conditions**. **a-e**, (I) The under-focused LTEM images showing skyrmions with the applied magnetic field at 0T (**a**), 0.5T (**b**), 0.7T (**c**) 0.8T (**d**), and the 0T field decreases from 0.8T (**e**), in which typical skyrmion is marked by a red ring. (II) Corresponding magnetic induction maps. **f-i**, (I) The under-focused LTEM images showing skyrmions with the applied temperature at 300K (**f**), 433K (**g**), 593K (**h**), and 673K (**i**). (II) Corresponding on-focused LTEM images, in which the PNRs's texture can be observed. **j**, The enlarged images of (**i**).

By applying a magnetic field perpendicular to the surface (Fig 2 (a-d)), the stability of skyrmions under an external magnetic field was further investigated. It can be observed that as the field increases from 0 T to 0.5 T, the skyrmions deform gradually from highly symmetric circles to ellipses. They demonstrated robust topological protection, preserving their topology structures until they reached a field strength of 0.7 T, surpassing the majority of previously reported magnetic skyrmions. Then, with the external field up to 0.8 T, the magnetic vortices were completely suppressed and transformed into striped magnetic domains. However, the skyrmions would gradually regenerate spontaneously when the external field was removed, indicating their remarkable non-volatile propertie (Fig 2 (e) & Supplement: Fig (7,8))

In order to evaluate the impact of temperature on excited skyrmions, we conducted high-temperature LTEM experiments to assess their stability. Fig 2 (f-i, I) presents over-focused LTEM images of the skyrmions at temperatures of 300 K (room temperature), 433 K, 593 K, and 673 K, respectively. The results show that the skyrmions maintain a remarkably stable vortex structure even at an elevated temperature of 673 K, higher than all previous results reported below 350 K[33,34], indicating the unparalleled high-temperature stability exhibited by LN-inspired skyrmions. Considering the ferroelectric environment induced by LN, it is imperative to verify the potential ferroelectric-related characteristics of magnetic skyrmions in this material. In order to eliminate any contrast response caused by magnetic signal, we adjusted the focal length of LTEM to 0 μm (also known as on-focused LTEM), to observe and analyze the potential ferroelectric traits of skyrmions. From the on-focused LTEM images (Fig. 2 (f-i, II)), we observed that a new grainy contrast signal appears in the center of the skyrmions at high temperatures, with the domains exhibiting a pattern of alternating dark and light stripes. This newly emerging signal indicates the existence of polar nanoregions (PNRs) within the central region of the skyrmions with a significant ferroelectric polarization[35,36]. Fig. 2 (j) respectively shows the over-focused and on-focused LTEM images of the LN film at 673 K in a narrow field of view at 200 $nm^2$. It can be confirmed that the skyrmion keeps stable while the signal of PNRs occurs. The size of the ferroelectric domain strips at the center of the skyrmion is approximately 6×20 $nm^2$ while the overall structure of PNRs exhibits a subcircular shape with a diameter close to 20 nm, indicating the localization character for the ferroelectricity of these skyrmions.

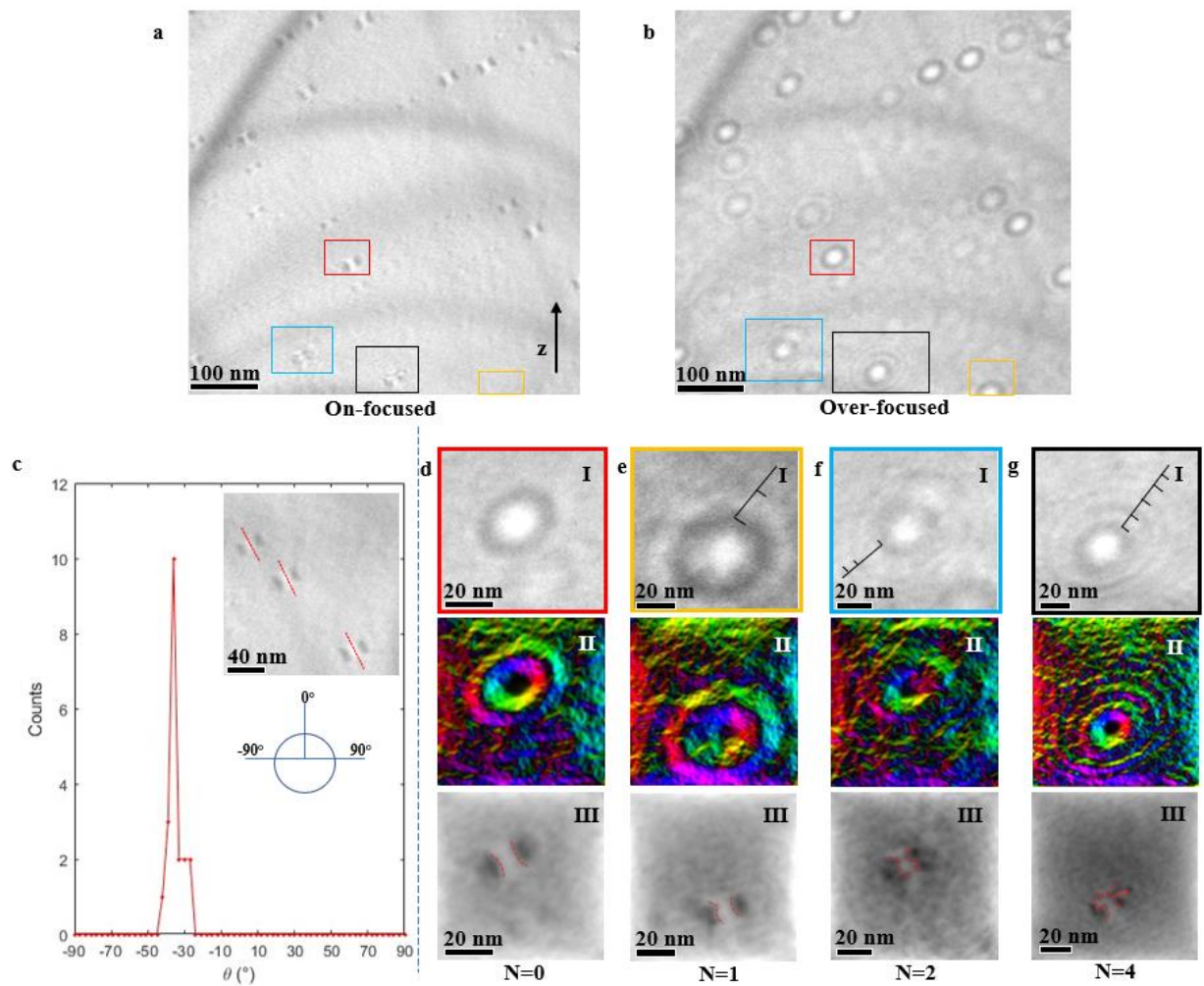

**Fig. 3 | PNRs's morphology with different types of skyrmions**. **a-b**, The on-focused and over-focused images within an expansive field of vision, in which the skyrmions with specific ring number are indicated by blocks. **c**, Statistical results about the domain wall orientation of PNRs. **d-g**, (I) The over-focused LTEM images showing individual skyrmion with N = 0 (**d**), N = 1 (**e**), N = 2 (**f**), and N = 4 (**g**), while the line segments are marked on each ring. (II) Corresponding magnetic induction maps. (III) Corresponding PNRs's maps, in which the ferroelectric domain boundaries are delineated by red dashed lines to facilitate resolution.

The LTEM images of the LN film, acquired at room temperature after cooling from 673 K, are presented in Fig 3 (a,b), illustrating the over-focused and on-focused states respectively. Fig 3 (a) shows that PNRs remained in room temperature, meaning that the ferroelectricity of the multiferroic skyrmions can remain stable without high temperature. It can also be observed that the PNRs of each multiferroic skyrmion possess a specific orientation and exhibit a relatively stable consistency around - 40°, concluded by measuring the angle between the direction of the domain walls of each skyrmion and the z-direction of LN (Fig 3 (c)). This phenomenon can be attributed to the anisotropy inherent in LN single crystals, considering the high symmetry in the magnetic vortex.

Furthermore, these images also unveil the emergence of novel magnetic vortex states within the multiferroic skyrmions, wherein certain skyrmions exhibit peripheral regions with multi-layer rings resembling hopfions[37]. Moreover, significant modifications in the textures of the PNRs can be observed within these skyrmions. These structures are magnified in Fig 3 (d-g), with the number of outer rings denoted by N. Here, dotted red lines indicate the positions of domain walls of PNRs, facilitating easy differentiation of the ferroelectric domain structures. In the case of a skyrmion with N = 0 and 1 (also known as skyrmion bundle[38]), the ferroelectric domains at the center of the vortex exhibit identical characteristics to those observed in LTEM at high temperatures, displaying alternating dark and light stripes. When N = 2, the bright fringe at the center of the ferroelectric domain extends from middle to outer regions, while the dark fringes on both sides are divided with the striped structure transforming into a point-like configuration. When N = 4, it can be observed that the central position of the bright stripe protrudes further outward, assuming a cross-like shape. Based on these observations, it is evident that changes in ferroelectric morphology occur with the alterations in magnetic vortex structures in multiferroic skyrmions. Considering the exclusive existence of these PNRs solely within the skyrmions and their confinement by the magnetic vortices, it can be inferred that potential interactions exist between the magnetic and ferroelectric properties of multiferroic skyrmions.

## Pockerls effect of the multiferroic skyrmions

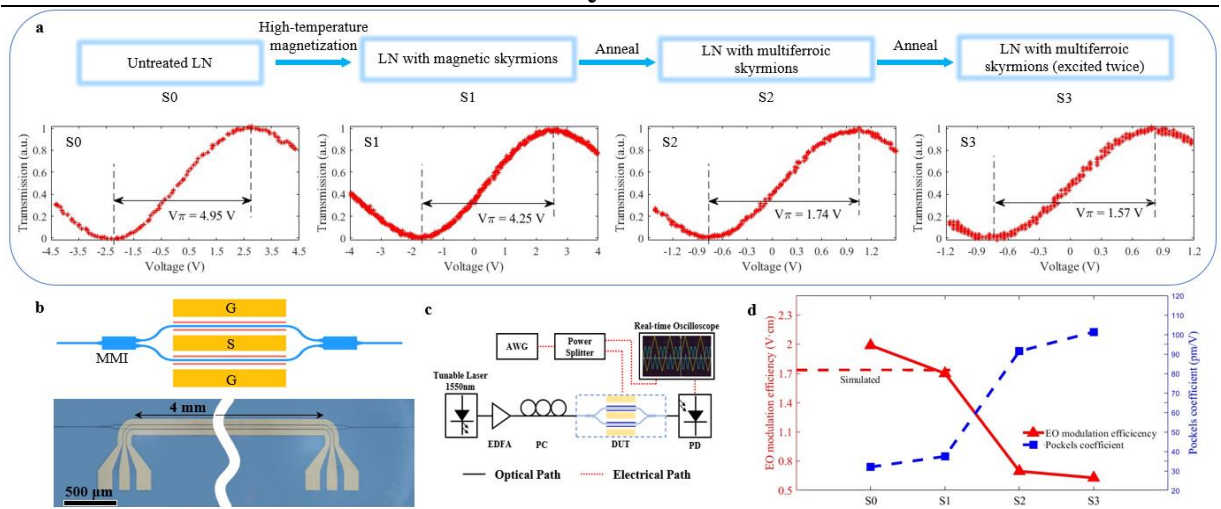

**Fig. 4** | **Results of EO modulator**. **a**, The half-wave voltage test results of the EO modulator after undergoing multi-stage processing. **b** The diagram depicts the structure of the EO modulator under test, while the microscope photograph showcases the fabricated device. The electrode positioned in the center of the device is the signal electrode (S), while the remaining electrodes on both sides serve as ground electrodes (G). **c**, Setup for the half-wave voltage measurement **d**.The statistical analysis of the EO modulation efficiency and the estimation of the Pockels coefficient respectively. The red dotted line represents the EO modulation efficiency obtained through simulation, reported in ref [39].

After confirming the ferroelectric properties of magnetic skyrmions in LN, we further explored the potential impact of this novel ferroelectric property. Given that the Pockels effect is one of the most widely used ferroelectric effects in LN materials, it naturally became our primary focus. Here, we employed a Mach-Zehnder structure-based EO modulator to assess the potency of the Pockels effect of LN with the multiferroic skyrmions (shown in Fig 4 (b)). The modulator utilizes the X-cut TFLN platform, while its device structure is based on previously published results[39]. In this experiment, the length of the modulation area is 4 mm, and the other design parameters are provided in Supplement: Fig 1 and Table 1.

The test setup can be found in Fig 4 (c) and Supplement: Sample Testing. The results are presented in Fig 4 (a), where the modulation efficiency of the device is evaluated at a frequency of 1 MHz with the optical wavelength at 1550nm. Without skyrmion excitation, the modulator exhibits a half-wave voltage ($V_\pi$) of 4.95 V, corresponding to a device modulation efficiency of 1.98 V·cm (S0). Then, the device underwent a magnetic field annealing process, resulting in the formation of skyrmions within LN (S1). Under these conditions, $V_\pi$ of the modulator decreases to 4.25 V; however, it remains within the expected theoretical range[39]. Subsequently, a rapid annealing process was performed on the device after magnetic field annealing, following the specific procedure outlined in Supplement: Sample Treatment. This annealing process resembles the heating procedure employed in high-temperature LTEM and facilitates the acquisition of ferroelectricity by skyrmions (S2). Remarkably, we observed e $V_\pi$ decrease from 4.25 V to 1.74 V, a value significantly lower than that predicted by simulations[43], indicating a significant enhancement in the modulation efficiency of the device. These findings suggest that the Pockels coefficient of LN with multiferroic skyrmions surpasses its initial value of 32 pm/V. The subsequent annealing iterations further diminish the half-wave voltage (S3). However, the potential for reduction at this stage is quite limited, resulting in a mere decrease of $V_\pi$ from 1.74 V to 1.57 V, thereby essentially approaching its threshold. The modulation efficiency and corresponding Pockels coefficient of the device throughout the process are illustrated in Figure 4 (d). In comparison to the initial state, when stimulated by multiferroic skyrmions, the device's modulation capacity increases from 1.98 V·cm to 0.63 V·cm, reaching a level that is 3.2 times as high as untreated samples. Moreover, the skyrmion state contributes to an exceptionally high Pockels coefficient of 69 pm/V, surpassing LN crystal itself and most current ferroelectric materials.

## Outlook

In summary, we employed a magnetic field annealing technique to excite room-temperature magnetic skyrmions in LN. This represents the first observation of the magnetic skyrmion state in non-magnetic materials, suggesting a potentially novel physical mechanism. The skyrmion exhibits exceptional physical properties, including a magnetic vortex structure with dimensions near 30 nm and remarkable stability against magnetic fields. The critical temperature of the skyrmion is over 673 K, surpassing all previously reported instances. Moreover, we found these magnetic skyrmions can be transformed to multiferroic skyrmions by means of a rapid annealing process and the PNRs occurred within the core of the magnetic vortex in the meantime. The multiferroic skyrmions can also maintain stability at room temperature, and some unique ones with multiple magnetic rings have been observed, indicating their distinctive physical properties.

The emergence of multiferroic skyrmion significantly enhances the EO coefficient of LN and greatly improves performance for the EO modulators based on TFLN platform with the modulation efficiency increased up to 3.2 times compared to original values. This discovery underscores the substantial value of SK-LN in the field of PICs and significantly broadens the potential applications of skyrmions. Furthermore, the cost-effective method we employ to generate skyrmions is fully compatible with CMOS technology, thereby endowing the SK-LN system not only with exceptional performance but also with complete industrial application capability. As a result, it holds considerable commercial prospects. Additionally, being the first non-magnetic system where skyrmions can exist and the only system where room-temperature multiferroic skyrmions can be observed up to now, LN offers a unique and distinctive platform research in the field of condensed matter physics and related areas involving multiferroic materials.

## Acknowledgements

We thank Miss Shuyan Shi, A.P. Zhong Li, and Suzhou Institute of Nano-Tech and Nano-Bionics, CAS and the East Changing Technologies for the provision of assistance in sample handling，A.P. Xiaolian Liu, Prof. Xuefeng Zhang，Prof. Guanghan Cao, and Miss. Jiayi Lu for the the provision of assistance in sample testing.We also appreciate the technical support from Mrs. Yanhua Liu and the ZJU Micro-Nano Fabrication Center, Zhejiang University.

Contributions

Y.Y. conceived the research and designed the experiments. T.C presided over and supported the research. N.C designed the mudulator. Y.Y. and Q.J.M fabricated the modulator. Y.Y., Y.R., N.C., K.L. and R.W. contributed to the measurements. Y.Y. and S.W. analysed the data. Y.Y., T.C., B.X. and N.C. wrote the manuscript.

Competing interests

The authors declare no competing interests.

# Supplement

**Device fabrication**

Devices are fabricated on a commercial X-cut LN-on-insulator wafer from NANOLN, which is widely adopted in research associated with LN. The bulk LN crystals are from the same company.

The device was fabricated using the CMOS process. The photolithography step involved the use of an I-line photolithography system (NIKON-I12) and an I-line photoresist (SPR-955). The $SiO_2$ layer was grown using Plasma Enhanced Chemical Vapor Deposition (PECVD: Oxford Plasmalab System 100). LN and $SiO_2$ etching were performed using Inductively Coupled Plasma (ICP: Oxford Plasmapro100 Cobra 180), while the gold layer growth was achieved through electron beam evaporation (EBE: ULVAC Cooke). The preparation process of the device is illustrated in Fig 1. Each step of the stepper process consists of two stages: homogenization and lithography. The remaining steps are carried out in conjunction with the stepper to transfer the adhesive pattern onto different layers of the device. The design parameters about modulation efficiency are depicted in Fig 2, while the corresponding values are enumerated in Table 1.

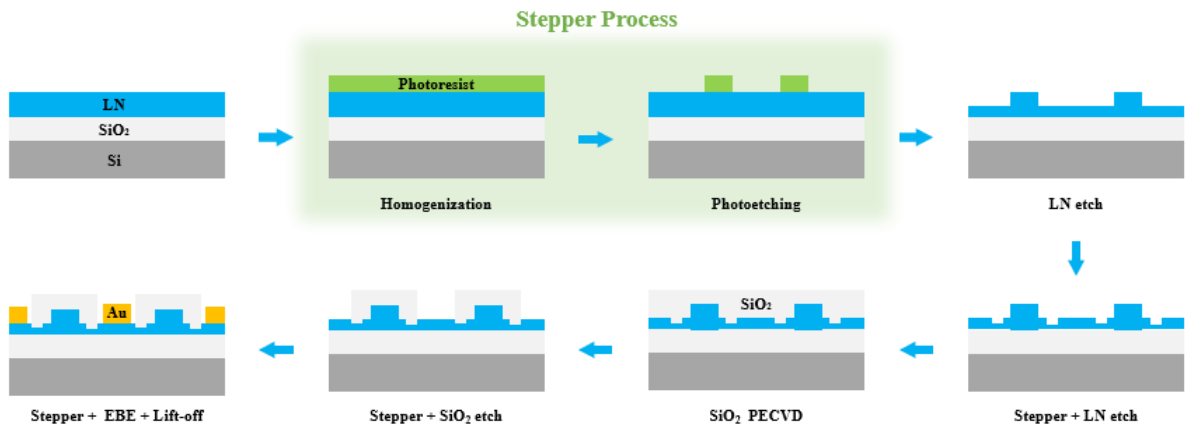

**Fig. 1** The processes for fabrication of EO modulator fabrication

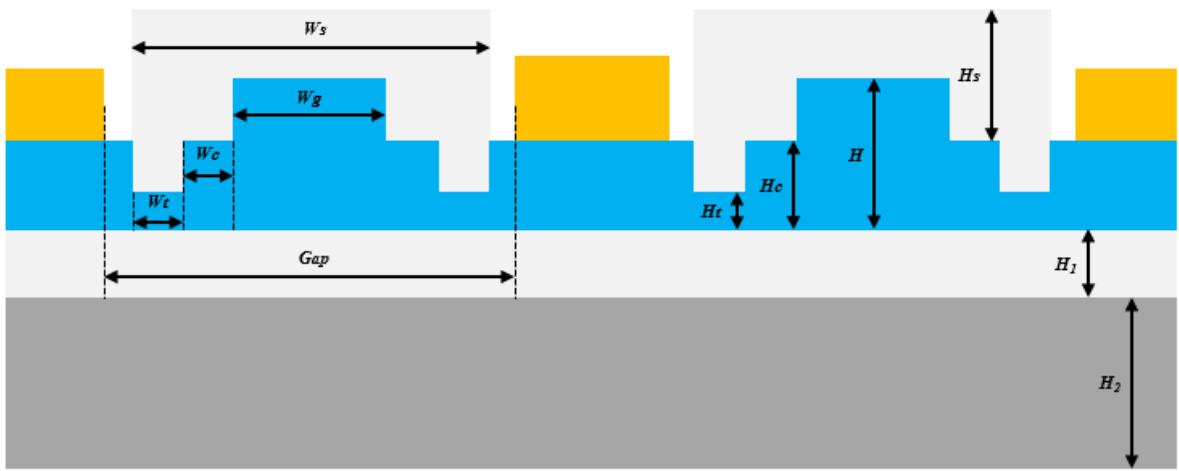

**Fig. 2** The schematic diagram depicts the modulation region of the device

| Designation | Value | Designation | Value |
| Wg | 1.4 μm | H | 0.6 μm |
| Ws | 3.2 μm | $H_1$ | 4.7 μm |
| Wc | 0.3 μm | $H_2$ | 550 μm |
| Wt | 0.5 μm | Hc | 0.4 μm |
| Gap | 3.6 μm | Ht | 0.2 μm |
| | | Hs | 0.8 μm |

**Table. 1** The design parameters of the EO modulator.

## Sample Treatment

We excite the skymions of lithium niobate through the following two steps: magnetic field annealing and rapid annealing. The magnetic field annealing process is conducted using the magnetization annealing furnace (East Changing technologies F800-35/EM7), while the rapid annealing process is achieved through the rapid annealing furnace (JouleYacht IRLA-1200).

1、   **Magnetic field annealing**

   The magnetic field is applied along the z direction of the LN, and the specific processing procedure is illustrated in Fig 3. The sample was exposed to a magnetic field of 1 Tesla at room temperature (290 K) and subsequently rapidly heated to 733 K at a rate of 100 K per minute. After a duration of 30 minutes, it was allowed to cool naturally, with the removal of the magnetic field occurring near 653 K, followed by further cooling down to 473 K. Subsequently, air injection facilitated rapid cooling back to room temperature.

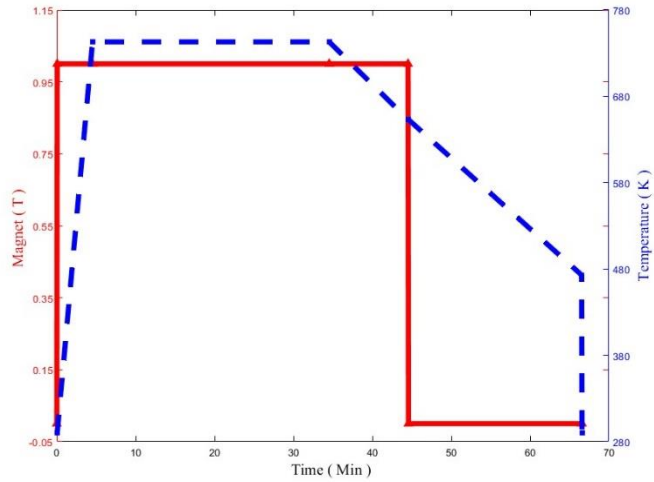

**Fig. 3** The procedure of magnetic field annealing

**2、　Rapid annealing**

The treatment process is illustrated in Fig 4. The sample undergoes rapid heating to 823 K at a rate of 500 K/min and is held at this temperature for a duration of 20 minutes. Subsequently, it naturally cools down to 423 K before being rapidly cooled to room temperature using forced air

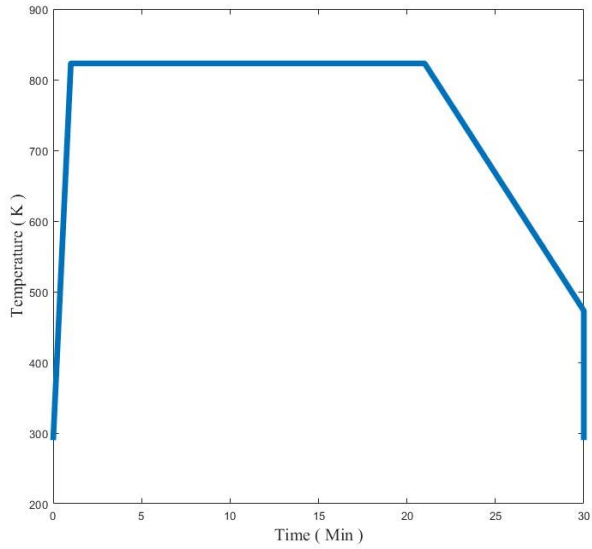

**Fig. 4** The procedure of rapid annealing.

## Sample Testing

### 1、 LTEM

The LTEM images were captured using an LTEM (ThermoFisher Talos F200S) system operated at 200 kV. The sample was sectioned into a thin film with a thickness of approximately 50 nm using the Focused Ion Beam technique (FIB: Helios G4 PFIB) and subsequently affixed onto the copper grid.

The LTEM results of the LN thin film without any treatment are presented in FIG. 5 (a-c), while FIG. 5 (d-f) displays the LTEM results of the LN thin film after magnetic field annealing, providing evidence that skyrmions are induced by the magnetic field annealing process.

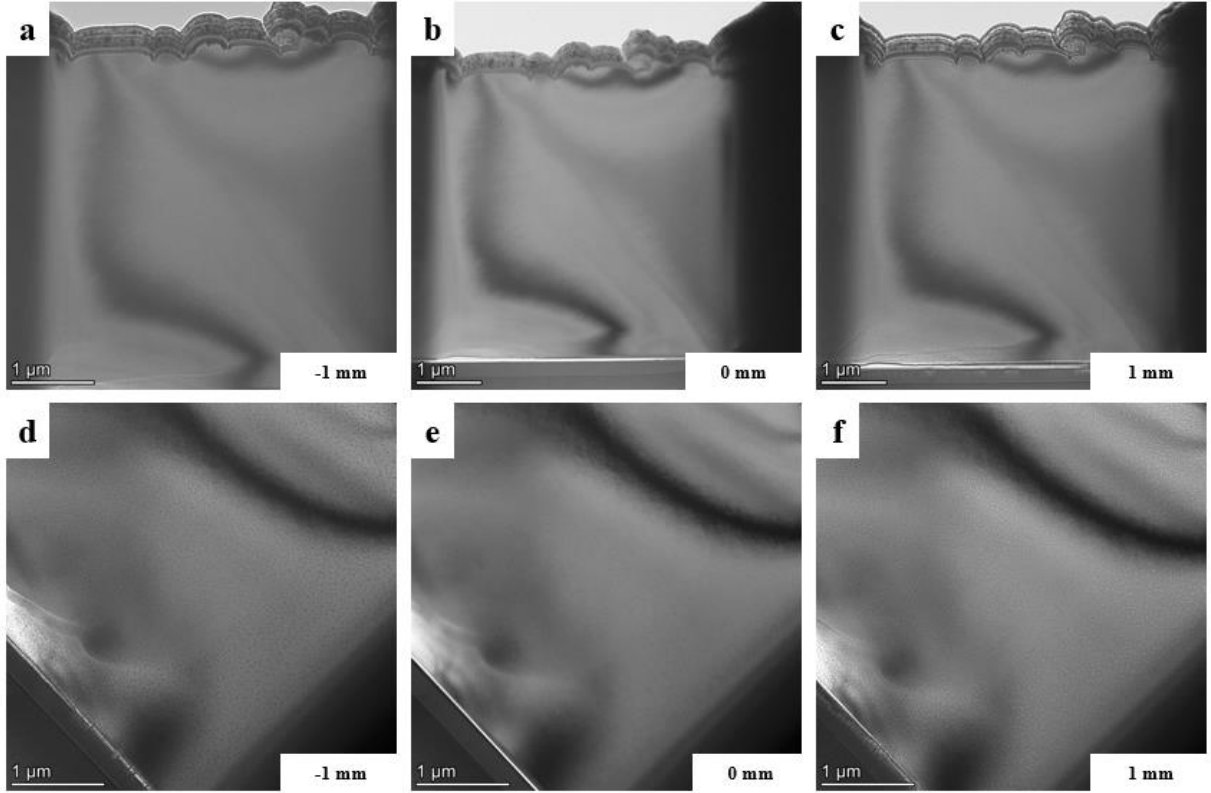

**Fig. 5** The LTEM images of LN. **a-c** Untreated. **d-f** After magnetic field annealing process.

2、 **Half-wave voltage measurement**

We characterized the modulation efficiency of the modulators using the experimental setup depicted in Figure 6. The light with a wavelength of 1550 nm emitted by a tunable laser (SANTEC TSL-550) was amplified by an Erbium-Doped Fiber Amplifier (EDFA: Keopsys CEFA-C-PB) and then coupled into the device under test (DUT) through a polarization controller (PC). The modulated light was detected using a high-speed photodetector (PD: LSIHPDA12G). An arbitrary signal generator (AWG: GW Instek AFG-3051) was utilized to provide a voltage sweep at 1 MHz. A real-time oscilloscope was employed to capture the signals from both the AWG and PD.

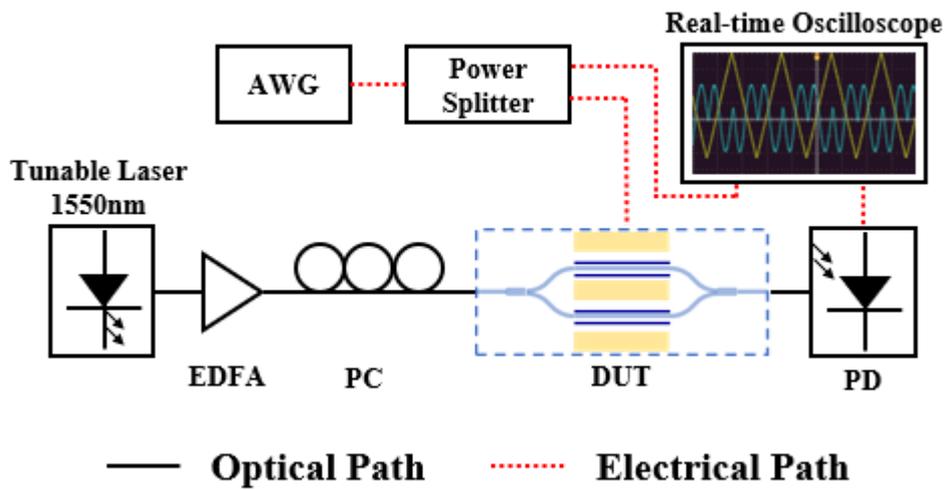

**Fig. 6** Setup for the half wave voltage measurement.

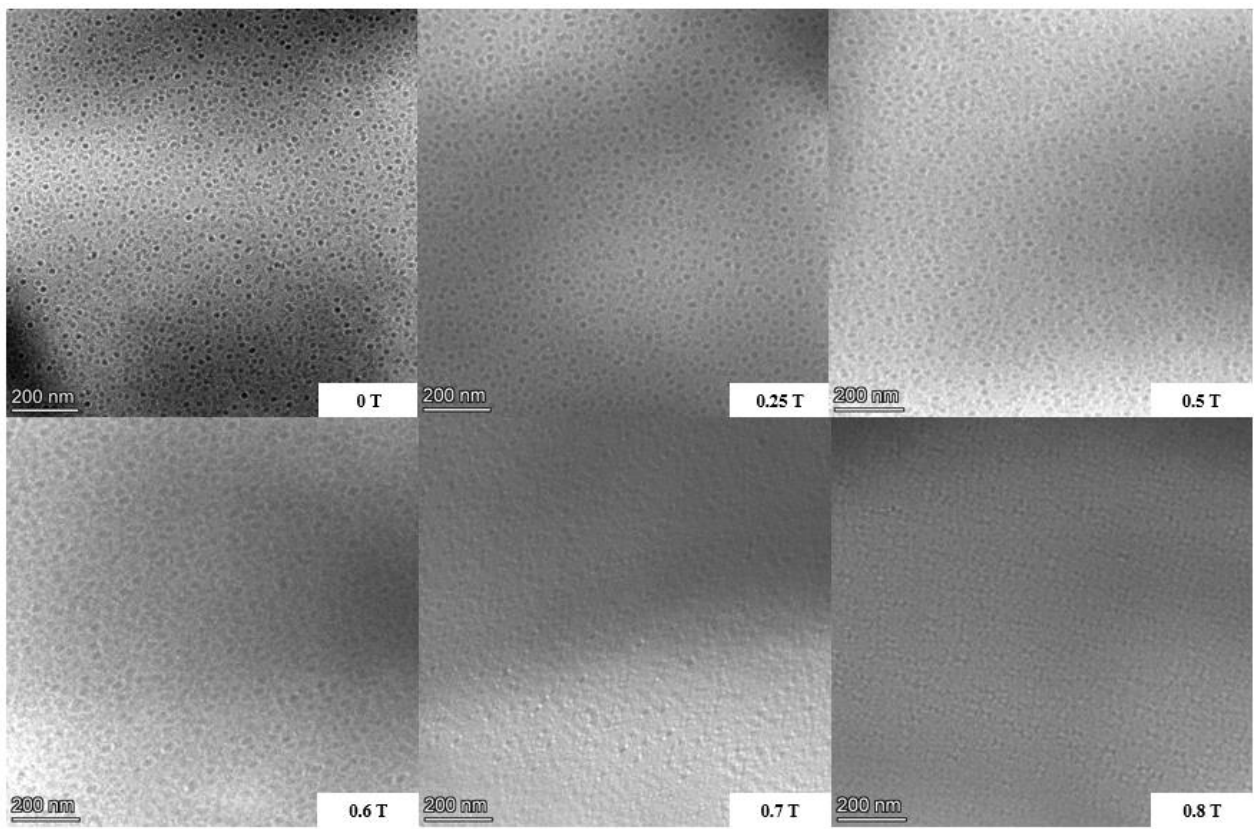

**Fig. 7** The LTEM images of skyrmions with field increased from 0 T to 0.8 T.

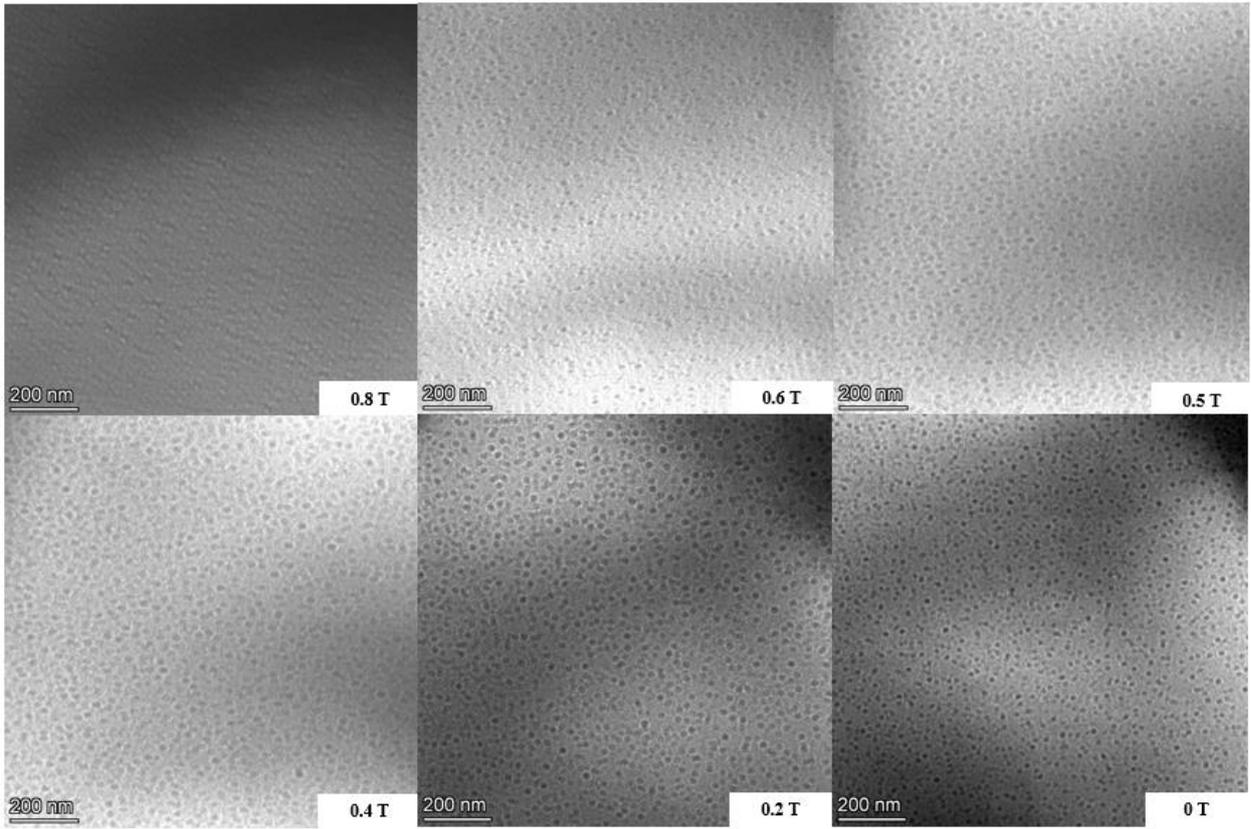

**Fig. 8** The LTEM images of skyrmions with field decreased from 0.8 T to 0 T.